\documentstyle{elsart}

\newfont{\bg}{cmr10 scaled\magstep4}
\makeatletter

\def\lsim{\compoundrel<\over\sim}
\def\compoundrel#1\over#2{\mathpalette\compoundreL{{#1}\over{#2}}}
\def\compoundreL#1#2{\compoundREL#1#2}
\def\compoundREL#1#2\over#3{\mathrel
      {\vcenter{\hbox{$\m@th\buildrel{#1#2}\over{#1#3}$}}}}
\makeatother

\begin{document}

\begin{frontmatter}

\title{Non-Adiabatic Transition in Spin-Boson Model
       and Generalization of the Landau--Zener Formula}
\author{Hiroto Kobayashi}\footnote{Present address: 
  c/o Prof.\ A.\ Kuniba, Institute of Physics, University of Tokyo,
  Komaba, Meguro-ku, Tokyo 153-8902, Japan; Tel.: (+81)-3-5454-4372;
  e-mail: hiroto@shpb.c.u-tokyo.ac.jp.}
\address{Department of Physics, Graduate School of Science, University
  of Tokyo \\ Hongo, Bunkyo-ku, Tokyo 113-0033, Japan}
\author{Naomichi Hatano}
\address{Theoretical Division, MS-B262, Los Alamos National
  Laboratory \\ Los Alamos, New Mexico 87545, USA}
\author{Seiji Miyashita}
\address{Department of Earth and Space Science, Faculty of Science,
  Osaka University \\ Toyonaka, Osaka 560-0043, Japan}

\begin{abstract}
Non-adiabatic transitions are studied in a spin-boson model with
multiple scattering points. In order to generalize the Landau--Zener
formula, which describes the case of a single scattering point, we
define an ``effective gap'' for a set of scattering points. The
generalized formula agrees very well with numerical results of the
non-adiabatic dynamics, which we obtained by a direct numerical
method. This will make the Landau--Zener formula yet more useful in
analyzing experimental data of magnetic-moment inversion.
\end{abstract}


\end{frontmatter}

\setcounter{equation}{0}
\section{Introduction}

Macroscopic quantum tunneling [1--4] 
has been an intriguing subject for several years. A few experiments
have been carried out \cite{aws1,tej} for mesoscopic magnetic systems
in order to determine the coupling matrix element (or the gap) between
the two macroscopic states. The basic assumption in analysis of the
experimental data is the system Hamiltonian of the form
\begin{equation}
{\cal H} =   \frac{1}{2} \hbar \delta \sigma^x
           - \frac{1}{2} H \sigma^z,            \label{eq:spin-model}
\end{equation}
where the up-spin state $ | \uparrow \rangle $ and the down-spin state
$ | \downarrow \rangle $ denote the two macroscopic states, while 
$ \delta (>0) $ in the first term denotes the coupling matrix element
between the two states. The coupling $ \hbar \delta $ is also called
the gap, the reason of which is obvious in the schematic energy
spectrum in Fig.\ \ref{fig:5.0}. 

The major experimental method of measuring the gap $ \hbar \delta $ is
to apply an oscillating field to the system and seek resonance due to
the gap at $ H = 0 $. However, this method may pick resonance of
various other degrees of freedom in real materials and mix them up
with the resonance due to the gap \cite{garg2}. Hence, one of the
present authors \cite{miyalz} suggested applying a constantly changing
field instead of an oscillatory field and estimating the coupling 
$ \hbar \delta $ by the use of the Landau--Zener formula
\cite{zen,landau}. 

Suppose that the system is first under a large field in the negative
direction $ ( H \ll - \hbar \delta ) $ and the spin is approximately
in the down-spin state. If the field is switched to the positive
direction very slowly, or adiabatically, the system follows the ground
state and eventually converges to the up-spin state for 
$ H \gg \hbar \delta $; thus the spin is flipped. If the field is
changed more rapidly, however, the system does not necessarily follow
the ground state and may remain in the down-spin branch even after the
field becomes positive. In other words, the system is excited to the
higher energy level at $ H = 0 $ with a certain probability $ p $. 
This phenomenon is called non-adiabatic transition. Hereafter, we
refer to the point where the transition takes place as the scattering
point. In this case, the scattering point is $ H = 0 $.

The Landau--Zener theory \cite{zen,landau} gives the ``passing''
probability $ p $ in the following form:
\begin{equation}
p = \exp \left[ - \frac{\pi}{2 \hbar} (\Delta E)^{2} v^{-1} \right]. 
\label{eq:LZ}
\end{equation}
Thus the spin is flipped with the probability $ 1 - p $. Here 
$ \Delta E $ is the gap at the scattering point; hence $ \Delta E = 
\hbar \delta $ in the case of Eq.\ (\ref{eq:spin-model}). The
parameter $ v $ denotes the rate of the field change; that is, the
field is increased from a large-magnitude negative value to a large
positive value as $ H(t) = vt $, where $ t $ is the time. Hence, to
measure the flipping probability under a constantly changing field
should give an estimate of the gap $ \hbar \delta $ from 
Eq.\ (\ref{eq:LZ}) \cite{miyalz}, provided that the above Hamiltonian
(\ref{eq:spin-model}) approximates experimental systems well.

In real materials, however, there must be always coupling between the
magnetic moment and phonons. Hence, in the present paper, we study
non-adiabatic transitions in the following spin-boson model
\cite{leg}: 
\begin{equation}
{\cal H} = - \frac{1}{2} \hbar \delta \sigma^x 
     - \frac{1}{2} H(t) \sigma^z 
     + \sum_\alpha \hbar \omega_\alpha 
        \left( n_\alpha + \frac{1}{2} \right) 
     + \sum_\alpha V_\alpha \sigma^z 
            ( b^\dagger_\alpha + b_\alpha ) .   \label{eq:sbhe}
\end{equation}
The third term denotes various phonon modes and the fourth term
denotes the coupling between the spin and the phonons. As we show
below, the energy spectrum of the model (\ref{eq:sbhe}) have many
scattering points at various values of $ H $. The main purpose of the
present paper is to show how we should modify the Landau--Zener theory
for such a problem. To generalize the formula (\ref{eq:LZ}) to
multiple-scattering cases, we introduce a new parameter, the effective
gap. The effective gap, when it replaces the gap $ \Delta E $ in the
Landau--Zener formula (\ref{eq:LZ}), explains well the
multiple-scattering data which we numerically calculated using the
exponential product formula \cite{fro,kob2}.

If the scattering points are well separated from each other in the
energy spectrum, a non-adiabatic transition at {\it each} scattering
point can be regarded as an independent event and should be well
approximated by the Landau--Zener formula (\ref{eq:LZ}). We can
explain the entire time evolution as a simple sequence of these
independent events. This is the case if the model (\ref{eq:sbhe}) has
only one boson mode. We first consider this case in Sec.\ 2 and show
that the Landau--Zener formula is indeed consistent with the numerical
results. 

However, this naive description should fail if some of the scattering
points are close to each other. This situation arises when we consider
multiple boson modes. We show in Sec.\ 3 that the numerical results in
two-mode cases can be well parametrized with a single parameter,
namely, the effective gap. 

For numerical calculations of the dynamics of the system, we changed
the magnetic field from $ - H_{\rm max} $ to $ H_{\rm max} $ in the
time range $ - t_{\rm max} \le t \le t_{\rm max} $; hence the rate of
the field change is $ v = H_{\rm max} / t_{\rm max} $ and the change
schedule is 
\begin{equation}
H(t) = vt = \frac{H_{\rm max}}{t_{\rm max}} \  t \quad {\rm for} \quad 
- t_{\rm max} \le t \le t_{\rm max}.  \label{eq:epst}
\end{equation}
We made the maximum field strength $ H_{\rm max} $ much greater than 
$ \hbar \delta $ to ensure that the numerical results well approximate 
the non-adiabatic dynamics with $ H_{\rm max} = \infty $. We
calculated the time evolution of the system, taking the ground state
for $ H = - H_{\rm max} $ as the initial state. Further computational
details are given in Appendix. 

Incidentally, we hereafter use dimensionless parameters for
simplicity, taking $ H_{\rm max} $ as the unit of energy. A tilde mark
on top of each parameter indicates that the parameter is
dimensionless: {\it e.g.}\ 
$\tilde{\cal H}\equiv{\cal H}/H_{\rm max}$,
$\tilde{\delta}\equiv\hbar\delta/H_{\rm max}$,
$\tilde{\omega}_{\alpha}\equiv\hbar\omega_{\alpha}/H_{\rm max}$,
$\tilde{V}_{\alpha}\equiv V_{\alpha}/H_{\rm max}$,
$\tilde{H}(\tilde{t})\equiv H(t)/H_{\rm max}$,
and $\tilde{t}\equiv tH_{\rm max}/\hbar$.
Hence the Hamiltonian (\ref{eq:sbhe}) now reads
\begin{equation}
\tilde{\cal H} = - \frac{1}{2} \tilde{\delta} \sigma^x 
     - \frac{1}{2} \tilde{H}(\tilde{t}) \sigma^z 
     + \sum_\alpha \tilde{\omega}_\alpha 
        \left( n_\alpha + \frac{1}{2} \right) 
     + \sum_\alpha \tilde{V}_\alpha \sigma^z 
            ( b^\dagger_\alpha + b_\alpha )    \label{eq:sbhe-dimless}
\end{equation}
with the field-change schedule given by
\begin{equation}
\tilde{H} (\tilde{t}) = \tilde{v} \tilde{t}
\quad \mbox{with} \quad
\tilde{v} = \frac{\hbar}{{H_{\rm max}}^{2}} \  v
          = \frac{1}{\tilde{t}_{\rm max}}
\quad \mbox{for} \quad 
- \tilde{t}_{\rm max} \le \tilde{t} \le \tilde{t}_{\rm max}.
\label{eq:epst-dimless}
\end{equation}
In this dimensionless notation, the maximum field is 
$ \tilde{H}_{\rm max} = 1 $. The Landau--Zener formula reads
\begin{equation}
p = \exp \left[ - \frac{1}{2} \pi (\Delta \tilde{E})^{2}
    \tilde{v}^{-1} \right]. 
\label{eq:LZ-dimless}
\end{equation}

\setcounter{equation}{0}
\section{Non-Adiabatic Transition and the Landau--Zener Formula}

In the present section, we treat the spin-boson model with a single
boson mode. We calculate time evolution from the initial state and
analyze it with the Landau--Zener formula. 

The energy spectrum of the Hamiltonian (\ref{eq:sbhe-dimless}) is
exemplified in Fig.\ \ref{fig:5.1}, showing its dependence on 
$ \tilde{H} $. (Note that the spectrum is calculated for each fixed
value of $ \tilde{H} $, not under a changing field 
$ \tilde{H}(\tilde{t}) $.) The values of the Hamiltonian parameters
are fixed to 
\begin{equation}
\tilde{\delta} = 0.05 , \quad
\tilde{\omega}_1 = 0.3  , \quad
\mbox{and} \quad
\tilde{V}_1 = \frac{0.1}{\sqrt{0.3}}
\label{eq:param-onemode}
\end{equation}
throughout this section. (The reason of the seemingly strange choice
of $\tilde{V}_{1}$ is that the spin-boson coupling 
$ \tilde{V}_{\alpha}$ often scales as 
$ 1 / \sqrt{\tilde{\omega}_{\alpha}} $.)

In Fig.\ \ref{fig:5.1}, the states with positive slopes have a large
component of the spin-down state, while those with negative slopes
have a large component of the spin-up state. The states with the same
spin but different energies differ in the approximate number of
excited bosons. (These interpretations are indicated in 
Fig.\ \ref{fig:5.1}.) The initial state, which is the ground state for
$ H = - H_{\rm max} $, or $ \tilde{H} = -1 $, is hence approximately a
spin-down state without any bosons. As $ \tilde{H} $ is increased at a
finite rate $ \tilde{v} $, the state evolves along the ground state of
the Hamiltonian before it encounters a nearly degenerate state at the
scattering point at $ \tilde{H} = 0 $. Non-adiabatic transitions occur
afterward. 

For example, the magnetization of the spin 
$ \langle \sigma^z \rangle_{\tilde{t}}$ and the total energy 
$ \tilde{E}(\tilde{t}) = \langle \tilde{\cal H} \rangle_{\tilde{t}} $
evolve as shown in Fig.\ \ref{fig:5.2} when the magnetic field is
increased as scheduled in Eq.\ (\ref{eq:epst-dimless}) with 
$ \tilde{v} = 1/500 $. Here the angle brackets $ \langle \cdots
\rangle_{\tilde{t}} $ denote the expectation value with respect to the
wave function at time $ \tilde{t} $: $ \langle \Psi(\tilde{t}) \bigm | 
\cdots \bigm | \Psi(\tilde{t}) \rangle $. (See Appendix for details of
the computation.) The initial state (approximately $ | \downarrow , 0
\rangle $) is scattered when it passes the scattering point at 
$ \tilde{H} = 0 $. The resulting state for $ 0 \lsim \tilde{t} \lsim
0.3 \  \tilde{t}_{\rm max} $ is a superposition of the ground state
(approximately $ | \uparrow , 0 \rangle $) and the first-excited state
(approximately $ | \downarrow , 0 \rangle $). Hence the magnetization
is not fully inverted at $ \tilde{H} = 0 $. (The oscillation of the
magnetization after the non-adiabatic transition is due to the spin
precession. This becomes evident when we see that the energy of the
spin does not change as rapidly as the magnetization does; see 
Fig.\ \ref{fig:5.3}.)

The above explanation is supported by the time-evolution data of the
eigenstate weights $ p_i $, where each $ p_i $ denotes the weight of
the $ (i-1) $th excited state in the evolving state 
$ \Psi(\tilde{t}) $. In other words, we define the weights by 
$ p_i (\tilde{t}) = | w_i(\tilde{t}) |^2 $, where we expand the
evolving state $ \Psi(\tilde{t}) $ with respect to the eigenfunctions
$ \{ \phi_i (\tilde{H}(\tilde{t})) \} $ of the Hamiltonian at time 
$ \tilde{t} $ as follows: 
\begin{equation}
\Psi(\tilde{t}) = 
\sum_i w_i (\tilde{t}) \phi_i (\tilde{H}(\tilde{t})) .
\end{equation}
The result of the calculation for the present case is shown in
Fig.\ \ref{fig:5.4}. The state $ \Psi(\tilde{t}) $ for $ 0 \lsim
\tilde{t} \lsim 0.3 \  \tilde{t}_{\rm max} $ is the superposition of
the two states $ | \uparrow , 0 \rangle $ and $ | \downarrow , 0
\rangle $, as is expected.

As the applied field is increased further from $ \tilde{H} = 0 $, the
component $ p_2 $ of the evolving state $ \Psi(\tilde{t}) $ undergoes
a subsequent non-adiabatic transition around $ \tilde{t} = 0.3 \ 
\tilde{t}_{\rm max} $ and spawns the component $ p_3 $. The component
$ p_2(\tilde{t}) $ for $ 0.3 \ \tilde{t}_{\rm max} \lsim \tilde{t}
\lsim 0.6 \ \tilde{t}_{\rm max} $ corresponds to 
$ | \uparrow , 1 \rangle $, a spin-up state with one boson emitted,
while the spawned component $ p_3(\tilde{t}) $ in the same time range
is approximately $ | \downarrow , 0 \rangle $. 
The component $ p_3 $ undergoes further non-adiabatic
transitions afterward. Thus the final state for $ \tilde{t} = \infty $
will be a superposition of spin-up states with various numbers of
bosons. 

In this one-mode case, the scattering points are far apart from each
other. We then expect that the whole event is regarded as a series of
independent non-adiabatic transitions, each of which can be treated by
the Landau--Zener formula (\ref{eq:LZ-dimless}). Hence the passing
probability after the transition at the scattering point $ \tilde{H} = 
0 $ should be 
\begin{equation}
p_2 (\tilde{t}) = \exp \left[ - \frac{1}{2} \pi 
(\Delta \tilde{E}_{1})^2 \tilde{v}^{-1} \right]
\quad \mbox{for} \quad 0 \lsim \tilde{t} \lsim 0.3 \  
\tilde{t}_{\rm max} ,
\label{eq:lz}
\end{equation}
while the probability that the state remains in the lower level is
$ p_1 = 1 - p_2 $. Here $ \Delta \tilde{E}_1 $ corresponds to the
energy gap between the ground and the first excited states at 
$ \tilde{H} = 0 $. Similarly for the subsequent non-adiabatic
transition around $ \tilde{t} = 0.3 \  \tilde{t}_{\rm max} $, the
passing probability is given by $ \exp [ - \frac{1}{2} \pi 
(\Delta \tilde{E}_2)^2 \tilde{v}^{-1} ] $, where $ \Delta 
\tilde{E}_2 $ is the energy gap between the first and second excited
states at the scattering point. The passing probability for the set of 
the two scattering points with the gaps $ \Delta \tilde{E}_1 $ and 
$ \Delta \tilde{E}_2 $ should be given by
\begin{eqnarray}
p_3 (\tilde{t}) & = & 
\exp \left[ - \frac{1}{2} \pi (\Delta \tilde{E}_1)^2 
                               \tilde{v}^{-1} \right]
\exp \left[ - \frac{1}{2} \pi (\Delta \tilde{E}_2)^2 
                               \tilde{v}^{-1} \right]
\nonumber \\ & = &  
\exp \left\{ - \frac{1}{2} \pi 
     \left[ (\Delta \tilde{E}_{1})^2 + (\Delta \tilde{E}_2)^2 \right]
                               \tilde{v}^{-1} \right\}
\nonumber \\ &   & 
\qquad \mbox{for} \quad 0.3 \  \tilde{t}_{\rm max} \lsim \tilde{t}
\lsim 0.6 \  \tilde{t}_{\rm max}.
\label{eq:p3}
\end{eqnarray}
On the other hand, the probability of the first excited component in
this time range is 
\begin{eqnarray}
p_2 (\tilde{t}) & = &
\exp \left[ - \frac{1}{2} \pi (\Delta \tilde{E}_1)^2 
                               \tilde{v}^{-1} \right]
\left\{ 1 - 
\exp \left[ - \frac{1}{2} \pi (\Delta \tilde{E}_2)^2 
                               \tilde{v}^{-1} \right]
\right\}
\nonumber \\ &    & 
\qquad \mbox{for} \quad 0.3 \  \tilde{t}_{\rm max} \lsim \tilde{t}
\lsim 0.6 \  \tilde{t}_{\rm max}.
\end{eqnarray}
We can continue the calculation for subsequent transitions.

Indeed, the above theoretical argument explains our numerical data of
the magnetization very well, as is shown in Fig.\ \ref{fig:5.5}. Thus
we can in turn estimate the energy gap of each scattering point from
the time dependence of the magnetization \cite{miyalz,der}. 

However, we cannot naively apply the Landau--Zener theory to cases
where some of the scattering points lie closely to each other. This is
the central issue of Sec.\ 3. Before we proceed to the next section, a
remark on Eq.\ (\ref{eq:p3}) is in order; the expression can be
formally regarded as if the system underwent one non-adiabatic
transition at a single scattering point with an ``effective gap'' 
$ \Delta \tilde{E}^{({\rm eff})} $ defined in
\begin{equation}
  \left( \Delta \tilde{E}^{({\rm eff})} \right)^2 
= \left( \Delta \tilde{E}_1 \right)^2 
+ \left( \Delta \tilde{E}_2 \right)^2 , 
\label{eq:eff0}
\end{equation}
as has been pointed out by Kayanuma \cite{kaya}. We show in the next
section that the passing probability can be written in the
Landau--Zener form with an effective gap even when scattering points
are close to each other, although the value of the effective gap can
be different from the one given in Eq.\ (\ref{eq:eff0}). 

\setcounter{equation}{0}
\section{Successive Non-Adiabatic Transitions and Effective Gap}

In the present section, we consider the spin-boson model with two
boson modes. We discuss the case where non-adiabatic transitions
cannot be regarded as independent events. We introduce the effective
gap to generalize the Landau--Zener formula to such a case. We study
how the magnitude of the effective gap depends on the distance between 
two scattering points. 

In this section we keep the first boson mode the same as in 
Eq.\ (\ref{eq:param-onemode}) and change the parameters of the second
boson mode as follows: 
\begin{enumerate}
\renewcommand{\labelenumi}{(\roman{enumi})}
\setlength{\itemsep}{0pt}
\item $ \tilde{\omega}_2 = 0.35 $ and 
      $ \tilde{V}_2 = 0.1 / \sqrt{0.35} $;
\item $ \tilde{\omega}_2 = 0.31 $ and 
      $ \tilde{V}_2 = 0.1 / \sqrt{0.31} $;
\item $ \tilde{\omega}_2 = \tilde{\omega}_{1}= 0.3 $ and 
      $ \tilde{V}_2 = \tilde{V}_{1}= 0.1 / \sqrt{0.3} $.
\end{enumerate}
We calculated the system dynamics, again taking the ground state for 
$ \tilde{H} = -1 $ as the initial condition.

In the case (i), there are two scattering points quite close but still
apart. We show that the passing probability for this case can be given 
by the product of two passing probabilities with the real values of
the energy gaps. 

This is not true in the case (ii), where the two scattering points are
much closer than in the case (i). However, we show numerically that
the same formula of the passing probability is still applicable,
although we have to replace the real values of the gaps with an
effective gap. (In fact, the real energy gaps are not clearly defined
in the case (ii) anymore.) The remarkable point is that the single
parameter of the effective gap determines the passing probabilities
for various field-change rates. This is a non-trivial observation.

In the case (iii), the two modes are degenerate and the two scattering
points in question are reduced to one point. Then the original
single-scattering Landau-Zener formula becomes applicable again, but
with a single real energy gap instead of two gaps. 

We then show in the last subsection that the effective gap estimated
in the case (ii) smoothly interpolates the real energy gaps in the
cases (i) and (iii). Thus we conclude that the effective gap is a very
useful concept in analyzing numerical and experimental data.

\subsection{Case (i)}

We show in Fig.\ \ref{fig:5.6} the energy spectrum of the spin-boson
model with two boson modes in the case (i): $ \tilde{\delta} = 0.05 $, 
$ \tilde{\omega}_1 = 0.3 $, $ \tilde{\omega}_2 = 0.35 $, 
$ \tilde{V}_1 = 0.1 / \sqrt{0.3} $, and 
$ \tilde{V}_2 = 0.1 / \sqrt{0.35} $. The time evolution starts with
the ground state at $ \tilde{H} = -1 $ and goes through the scattering
points near $ \tilde{H} = 0 $, 0.3, 0.35,\ldots. We calculated the
time evolution of the magnetization $ \langle \sigma^z
\rangle_{\tilde{t}} $ of the spin and the total energy, as shown in
Fig.\ \ref{fig:5.7}. 

Let us focus on the scattering points near $ \tilde{H} = 0.3 $ and
0.35. Although these two are fairly separated, we cannot resolve the
magnetization jump around $ \tilde{H} = 0.3 \sim 0.35 $ into two
non-adiabatic transitions. The spin-precession oscillation probably
smeared out the two magnetization jumps. Still, we can resolve it into
two in the calculations of the eigenstate weights, the same quantities
as plotted in Fig.\ \ref{fig:5.4}. In Fig.\ \ref{fig:5.8}, we can
separate the two ``scattering regions,'' where the eigenstate weights
$p_{i}$ drastically change. Indeed, the values of the eigenstate
weights are consistent with the Landau--Zener formulas separately
applied to the two scattering points; i.e.\ the eigenstate weight 
$ p_2 $ converges to
\begin{equation}
p_2 =
\exp \left[ - \frac{1}{2} \pi (\Delta \tilde{E_{1}})^{2}
                               \tilde{v}^{-1} \right]
\left\{ 1 - 
\exp \left[ - \frac{1}{2} \pi (\Delta \tilde{E_{2}})^{2}
                               \tilde{v}^{-1} \right]
\right\}
\end{equation}
and $ p_3 $ to
\begin{eqnarray}
p_3 & = & 
\exp \left[ - \frac{1}{2} \pi (\Delta \tilde{E_{1}})^{2}
                               \tilde{v}^{-1} \right]
\exp \left[ - \frac{1}{2} \pi (\Delta \tilde{E_{2}})^{2}
                               \tilde{v}^{-1} \right]
\nonumber \\ &   & \times
\left\{ 1 -
\exp \left[ - \frac{1}{2} \pi (\Delta \tilde{E_{3}})^{2}
                               \tilde{v}^{-1} \right]
\right\}
\end{eqnarray}
for $ \tilde{t} = \tilde{t}_{\rm max} $. Here 
$ \Delta \tilde{E_{1}} $, $ \Delta \tilde{E_{2}} $, and 
$ \Delta \tilde{E_{3}} $ are the energy gaps at the scattering points
near $ \tilde{H} = 0 $, 0.3, and 0.35, respectively.

Since the magnetization behaves as if there is a single non-adiabatic
transition around $ \tilde{H} = 0.3 \sim 0.35 $, the situation
remarked at the end of the previous section now becomes a reality in
terms of the magnetization. That is, the magnetization jump around 
$ \tilde{H} = 0.3 \sim 0.35 $ is approximately described by a single
Landau--Zener formula (\ref{eq:LZ-dimless}) with the effective gap
\begin{equation}
       \left( \Delta \tilde{E}^{({\rm eff})} \right)^2 
\simeq \left( \Delta \tilde{E}_2 \right)^2 
     + \left( \Delta \tilde{E}_3 \right)^2 .
\label{eq:eff-case1}
\end{equation}
This is plotted in Fig.\ \ref{fig:5.7} as a dotted line with steps. 

\subsection{Case (ii)}

If two scattering points lie close within the width of the scattering
region, we cannot regard two non-adiabatic transitions as independent
events anymore. In the energy spectrum in the case (ii), the
scattering points near $ \tilde{H} = 0.3 $ and 0.31 are so close that
we cannot even define the energy gaps $ \Delta \tilde{E}_2 $ and 
$ \Delta \tilde{E}_3 $; see Fig.\ \ref{fig:5.9}. The time evolution of
the eigenstate weights $ p_i $ are shown in Fig.\ \ref{fig:5.10}. We
see from this figure that the two non-adiabatic processes are not
separate enough to be independent. 

In this case, we propose a new definition of the effective gap for the
non-adiabatic transitions around $ \tilde{H} = 0.3 \sim 0.31 $. With
this newly defined effective gap $ \Delta \tilde{E}^{({\rm eff})} $,
the magnetization jump around $ \tilde{H} = 0.3 \sim 0.31 $ is
described by a single Landau--Zener formula; see Fig.\ \ref{fig:5.11}.

Let us describe the definition of the effective gap in the
following. First, we define the effective gap 
$ \Delta \tilde{E}_{i}^{({\rm eff})} $ for each scattering point with
\begin{equation}
  p_{i} (\tilde{t}_{\rm max})
= \left[ 1 - \sum_{j=1}^{i-1} p_j (\tilde{t}_{\rm max}) \right]
  \left\{ 1 - 
\exp \left[ - \frac{1}{2} \pi (\Delta \tilde{E}_i^{({\rm eff})} )^2 
                               \tilde{v}^{-1}
    \right]   \right\} .
\label{eq:def-eff}
\end{equation}
Specifically, we obtain the first estimate 
$ \Delta \tilde{E}_1^{({\rm eff})} $ from the value of 
$ p_1 (\tilde{t}_{\rm max}) $ as
\begin{equation}
\Delta \tilde{E}^{({\rm eff})}_1 = 
\sqrt{ - \frac{ 2 \tilde{v} }{\pi} 
       \log [ 1 - p_1 (\tilde{t}_{\rm max}) ] } . 
\end{equation}
Since the scattering point at $ \tilde{H} = 0 $ is far apart from the
other scattering points, the estimate 
$ \Delta \tilde{E}_1^{({\rm eff})} $ simply coincides with the real
gap $ \Delta \tilde{E}_1 $ between the ground and first excited states
at $ \tilde{H} = 0 $. The second estimate 
$ \Delta \tilde{E}_2^{({\rm eff})} $ is given by
\begin{equation}
  p_2 (\tilde{t}_{\rm max})
= \left[ 1 - p_1 (\tilde{t}_{\rm max}) \right]
  \left\{ 1 - 
\exp \left[ - \frac{1}{2} \pi (\Delta \tilde{E}_2^{({\rm eff})} )^2 
                               \tilde{v}^{-1}
    \right]   \right\} ,
\end{equation}
or
\begin{equation}
\Delta \tilde{E}^{({\rm eff})}_2 = 
\sqrt{ - \frac{ 2 \tilde{v} }{\pi}
       \log \left[ 1 - 
  \frac{p_2 (\tilde{t}_{\rm max})}
   {1 - p_1 (\tilde{t}_{\rm max})} \right] } . 
\end{equation}
This effective gap, however, has no counterpart to be compared with,
since we cannot properly define the energy gap between the first and
second excited states at the scattering point near $ \tilde{H} = 
0.3 $. This is also the case for the third estimate 
$ \Delta \tilde{E}_3^{({\rm eff})} $. By adding the last two effective
gaps squared, we then define the combined effective gap 
$ \Delta \tilde{E}^{({\rm eff})} $ for the transitions around 
$ \tilde{H} = 0.3 \sim 0.31 $:
\begin{equation}
  \left( \Delta \tilde{E}  ^{({\rm eff})} \right)^{2}
= \left( \Delta \tilde{E}_2^{({\rm eff})} \right)^{2}
+ \left( \Delta \tilde{E}_3^{({\rm eff})} \right)^{2} .
\label{eq:gap-combined}
\end{equation}
The dotted line with steps in Fig.\ \ref{fig:5.11} was drawn, using
the Landau--Zener formula with the above combined effective gap. 

The key point here is that thus-defined effective gaps barely depend
on the field-change rate $ \tilde{v} $; see Fig.\ \ref{fig:5.11.5}.
(Otherwise the definition of the effective gaps would be just a
transformation of the eigenstate weights.) Therefore, once we estimate
the effective gaps for one particular value of $ \tilde{v} $, we can
predict the magnetization jump for arbitrary values of $ \tilde{v} $.
This is the central point of the present paper.

\subsection{Case (iii)}

The energy spectrum in the case (iii), $ \omega_1 = \omega_2 = 0.3 $,
is shown in Fig.\ \ref{fig:5.12}. The two scattering points around 
$ \tilde{H} = 0.3 $ are now reduced to one. In this case, the second
excited state at the scattering point near $ \tilde{H} = 0.3 $ is not
coupled to either the first or the third excited states
\cite{kaya}. This is demonstrated in Fig.\ \ref{fig:5.13}; the
component of the second excited state, $ p_3 $, entirely
vanishes. Since the second excited state becomes irrelevant near 
$ \tilde{H} = 0.3 $, the non-adiabatic transition there occurs only
between the first and the third excited states. Thus we recover the
original Landau--Zener problem for two states. The time dependence of
the magnetization is well explained by the Landau--Zener formula with
the irrelevant states neglected. The magnetization jump around 
$ \tilde{H} = 0.3 $ is given by the formula (\ref{eq:LZ-dimless}) with
the real energy gap between the first and the third excited states
(the dotted line with steps in Fig.\ \ref{fig:5.14}). 

\subsection{Effective gap and separation of two scattering points}

We have shown above that the case (i) and the case (iii) are explained 
by the original Landau--Zener theory for a single non-adiabatic
transition between two states. The passing probabilities are fully
described by the formula (\ref{eq:LZ-dimless}) with the real values of
the energy gaps at the scattering points. In the intermediate region
including the case (ii), on the other hand, we need to use the
effective gap for the formula, since the real energy gaps are not
clearly defined. We here show that the effective gap estimated in the
intermediate region smoothly interpolates the real energy gaps in the
limiting cases (i) and (iii).

We calculated the effective gap in the intermediate region, fixing 
$ \tilde{\omega}_1 = 0.3 $ and changing $ \tilde{\omega}_2 $ in the
range $ 0.3 \le \tilde{\omega}_2 \le 0.35 $. (We changed 
$ \tilde{V}_2 $ at the same time so that $ \tilde{V}_2 \propto 
1 / \sqrt{\tilde{\omega}_2}$.) By varying the value of 
$ \tilde{\omega}_2 $, we change the separation of the two scattering
points around $ \tilde{H} \sim 0.3 $. Figure \ref{fig:5.15} shows the
$ \tilde{\omega}_2 $-dependence of the effective gaps 
$ \Delta \tilde{E}^{({\rm eff})}_1 $, 
$ \Delta \tilde{E}^{({\rm eff})}_2 $, and 
$ \Delta \tilde{E}^{({\rm eff})}_3 $, which are defined in 
Eq.\ (\ref{eq:def-eff}). We can see that the effective gaps in the
intermediate region smoothly interpolate the two cases (i) and
(iii). This shows that the data analysis with the effective gap is
very stable. 

To put the above in another perspective, we show in 
Fig.\ \ref{fig:5.17} the discrepancy between the real energy gap and
the effective energy gap. In the case (i), the effective gap 
$ \Delta \tilde{E}^{({\rm eff})} $ is approximately given by the real
energy gaps as in Eq.\ (\ref{eq:eff-case1}), or
\begin{equation}
       \left( \Delta \tilde{E}^{({\rm eff})} \right)^2 
\simeq \left( \Delta \tilde{E}_2 \right)^2 
     + \left( \Delta \tilde{E}_3 \right)^2 .
\label{eq:eff-case1-again}
\end{equation}
We can see in Fig.\ \ref{fig:5.17} that the discrepancy at 
$ \tilde{\omega}_2 = 0.35 $ is only about 2\%. Although the real
energy gaps may not be clearly defined in the intermediate region 
$ 0.3 \le \tilde{\omega}_2 \le 0.35 $, let us, only for illustration
purposes, define (a) $ \Delta \tilde{E}_2 $ as the energy difference
between the first and the second excited states at 
$ \tilde{H} = \tilde{\omega}_1 = 0.3 $ and (b) $ \Delta \tilde{E}_3 $
as the energy difference between the second and the third excited
states at $ \tilde{H} = \tilde{\omega}_2 $. The right-hand side of 
Eq.\ (\ref{eq:eff-case1-again}) thus estimated (the dashed line in 
Fig.\ \ref{fig:5.17}) becomes apart from the effective gap when the
two scattering points get close to each other as $ \tilde{\omega}_2
\to \tilde{\omega}_1 $. Note that it is the effective gap, not the
real gaps, that describes the dynamics correctly. 

In the case (iii), the real gaps tentatively defined as above is now
related to the effective energy gap as 
$ \Delta \tilde{E}^{({\rm eff})} = \Delta \tilde{E}_2 + 
\Delta \tilde{E}_3 $. The left-hand side squared is the dotted line in
Fig.\ \ref{fig:5.17}. This becomes apart from the effective gap when
the two scattering points get separated from each other. Figure
\ref{fig:5.17} also shows that the effective gap is a useful way of
interpolating the two limiting cases. 

\setcounter{equation}{0}
\section{Summary}

In the present paper, we studied the time evolution of a spin-boson
model which may be relevant to systems with macroscopic quantum
tunneling. We introduced the definition of the effective gap to
explain multiple non-adiabatic transitions in complicated energy
spectra. With the use of the effective gap, the Landau--Zener formula,
originally obtained for a single scattering point, explains the
model's multiple magnetization jumps. The important point is that the
effective gap estimated in one case of time evolution (i.e.\ for
one value of $ \tilde{v} $) predicts the dynamics for all other values
of $ \tilde{v} $. This fact will make the Landau--Zener formula yet
more useful in analyzing experimental data of magnetic-moment
inversion. 

We remark on a related work on multiple non-adiabatic
transitions. Kayanuma and Fukuchi \cite{kaya} studied successive
non-adiabatic transitions in the following generalized case which
consists of assembly of the Landau--Zener scattering points: 
\begin{equation}
{\cal H} (t) = \left( \begin{array}{ccccc}
  v t  &    J_1    & J_2 & J_3 & \cdots    \\
  J_1  &    \varepsilon_1    &     &     & 
\smash{\lower1.7ex\hbox{\bg 0}} \\
  J_2  &           & \varepsilon_2 &     &           \\
  J_3  &           &     & \varepsilon_3 &           \\
\vdots & \smash{\hbox{\bg 0}} &     &     & \ddots  
\end{array} \right). \label{eq:kayahami}
\end{equation}
Here the parameters $ v $, $ \{ \varepsilon_i \} $, and $ \{ J_i \} $
are arbitrary constants. The theory concludes that the overall passing
probability is given by 
\begin{equation}
P = \exp \left( - \frac{2 \pi}{\hbar} \sum_i {J_i}^2 |v|^{-1} \right)
    \label{eq:kayalz}
\end{equation}
for arbitrary values of $ \{ \varepsilon_i \} $.

The energy spectrum of the above Hamiltonian appears quite similar to
the energy spectra of the present spin-boson model, particularly when
$ \frac{1}{2} vt $ is subtracted from all the diagonal elements of the
Hamiltonian (\ref{eq:kayahami}). In fact, consider the case where all
the off-diagonal couplings $ \{ J_i \} $ are much smaller than the
level spacings of $ \{ \varepsilon_i \} $. The structure of the energy
spectrum is almost equivalent to the energy spectrum in 
Fig.\ \ref{fig:5.1} except for the levels 
$ | \downarrow , n > 0 \rangle $ (which do not contribute to the
dynamics in any case). Then the correspondence $ \Delta E_i = 2 J_i $
becomes evident. 

To put it in other words, comparison of our conclusion with 
Eq.\ (\ref{eq:kayalz}) leads us to the speculation that our spin-boson
model may be reduced to an effective Hamiltonian of the form 
\begin{equation}
{\cal H}^{({\rm eff})} = \left( \begin{array}{cccc}
H(t)  & 
\frac{1}{2} \Delta E^{({\rm eff})}_1 & 
\frac{1}{2} \Delta E^{({\rm eff})}_2 & \cdots                      \\
\frac{1}{2} \Delta E^{({\rm eff})}_1 & \varepsilon_1        &    & 
                                   \smash{\lower1.7ex\hbox{\bg 0}} \\
\frac{1}{2} \Delta E^{({\rm eff})}_2 &           & \varepsilon_2 & \\
            \vdots                   & \smash{\hbox{\bg 0}} &    & 
                                                               \ddots 
\end{array} \right) - \frac{1}{2} H(t) . \label{eq:hamieff}
\end{equation}
However, the reduction procedure, if any, is yet to be known.

Finally, the exponential product formula reviewed in Appendix has
proven very useful in calculating non-adiabatic dynamics
accurately. We hope the formula plays an important role for many
problems of quantum dynamics in systems with more complicated
interactions. 

\appendix

\setcounter{equation}{0}
\section{Exponential Product Formula for Time-Dependent Hamiltonians}

In this Appendix we review the computational method employed in the
present paper, namely the exponential product formula \cite{fro}.

In our previous study \cite{kob2}, we discussed the time evolution of
the spin-boson model (\ref{eq:sbhe}) in a constant magnetic field
using the exponential product formula. In this case, the
diagonalization, if possible, of the Hamiltonian might suffice to know
the dynamics. In non-adiabatic transitions, on the other hand,
diagonalization of the Hamiltonian does not provide enough information 
for computation of its quantum dynamics, because of the time
dependence of the Hamiltonian. Since quantities change dramatically
when the system undergoes non-adiabatic transition, high precision is
required to calculations of the time evolution; even small
computational errors near the transition points can result in large
errors at later stages of the time evolution. The exponential product
formula is indispensable in such problems. 

When the Hamiltonian depends on time explicitly, a formal solution of
the Schr\"{o}dinger equation
\begin{equation}
i \hbar \frac{\partial}{\partial t} \Psi (t) = {\cal H} (t) \Psi (t)
\end{equation}
is given by $ \Psi (t) = U(t,0) \Psi (0) $, where $U(t,0)$ is the 
following time-ordered exponential:
\begin{eqnarray}
&   & U(t,0)                                           \nonumber \\
& = &          {\rm T} \exp \int_0^t 
 \left( - \frac{i}{\hbar}   \right)       {\cal H} (s  ) ds 
                                                       \nonumber \\
& = & 1 - \frac{i}{\hbar}   \int_0^t ds_1 {\cal H} (s_1)
        - \frac{1}{\hbar^2} \int_0^t ds_1 
                        \int_0^{s_1} ds_2 {\cal H} (s_1) 
                                          {\cal H} (s_2)
+ \cdots .                                             \nonumber \\
\end{eqnarray}
When we break up the present spin-boson Hamiltonian into three parts
as 
\begin{equation}
  {\cal H} (t) 
= {\cal H}_{\rm S} (t) + {\cal H}_{\rm B} + {\cal H}_{\rm I}
\end{equation}
with
\[
{\cal H}_{\rm S} (t) = - \frac{1}{2} \hbar \delta \sigma^x 
         - \frac{1}{2} H (t)  \sigma^z ,
\quad
{\cal H}_{\rm B} = \sum_\alpha \hbar \omega_\alpha 
          \left( n_\alpha + \frac{1}{2} \right) ,
\]
\begin{equation}
{\cal H}_{\rm I} = \sum_\alpha V_\alpha \sigma^z 
            ( b^\dagger_\alpha + b_\alpha ) ,
\end{equation}
the second-order approximant $ U_2 (t + \Delta t , t) $ to 
$ U (t + \Delta t , t) $ for a small time step $ \Delta t $ is 
given \cite{fro} by 
\begin{eqnarray}
&   & U_2 (t + \Delta t , t)               \nonumber \\
& = & \exp \left[ - \frac{i \Delta t}{2 \hbar}
     {\cal H}_{\rm S} \left( t + \frac{1}{2} \Delta t \right) \right]
      \exp \left( - \frac{i \Delta t}{2 \hbar}  
     {\cal H}_{\rm B} \right)
      \exp \left( - \frac{i \Delta t}  {\hbar} 
     {\cal H}_{\rm I} \right)
\nonumber \\ & & \times
      \exp \left( - \frac{i \Delta t}{2 \hbar}
     {\cal H}_{\rm B} \right)
      \exp \left[ - \frac{i \Delta t}{2 \hbar}  
     {\cal H}_{\rm S} \left( t + \frac{1}{2} \Delta t \right) \right] , 
\end{eqnarray}
i.e.,
\begin{equation}
U (t + \Delta t , t) = U_2 (t + \Delta t , t) + O((\Delta t)^2) . 
\end{equation}
Note that the partial Hamiltonian $ {\cal H}_{\rm S} $ is evaluated at
time $ t + \frac{1}{2} \Delta t $; this is essential in constructing
the second-order approximant to the time-ordered exponential
\cite{fro}. Matrix elements of the approximate time-evolution operator
are calculated in the same way as in the previous paper \cite{kob2}.

\newpage

\renewcommand{\thefigure}{\arabic{figure}}

\section*{Figure captions}

Figure \ref{fig:5.0}: A schematic of the energy spectrum of the
two-level system (\ref{eq:spin-model}) (the solid line). The dashed
lines denote the eigenvalues of the second term only. The labels 
$ | \uparrow \rangle $ and $ | \downarrow \rangle $ indicate the
states in the limit $ H \to \pm \infty $.

Figure \ref{fig:5.1}: The energy spectrum in the one-mode case 
  (\protect\ref{eq:param-onemode}). The abscissa is the magnetic field 
  $ \tilde{H} $. The labels $ | \downarrow , 0 \rangle $ and
  $ | \uparrow , n \rangle $ indicate the relevant physical
  interpretation of the states, where $ n $ is the number of excited
  bosons.

Figure \ref{fig:5.2}: The time evolution of the magnetization 
  $ \langle \sigma^z \rangle_{\tilde{t}}$ (dotted line) and the total
  energy $ \tilde{E}(\tilde{t}) $ (dashed line). The field was changed
  as $ \tilde{H} = (1/500) \tilde{t}$. The ground-state energy for
  each value of $ \tilde{H} $ is also shown as the solid line. 

Figure \ref{fig:5.3}: The time evolution of the magnetization 
  $ \langle \sigma^z \rangle_{\tilde{t}} $ (dashed line) and the
  energy of the spin, $ \tilde{E}_{\rm spin} (\tilde{t}) \equiv 
  \langle - \frac{1}{2} \tilde{\delta} \sigma^x 
          - \frac{1}{2} \tilde{H}(\tilde{t}) \sigma^z 
  \rangle_{\tilde{t}}$ (solid line).

Figure \ref{fig:5.4}: The time evolution of the eigenstate weights 
  $ p_i $. The solid line $ p_1 $ indicates the component of the
  ground state. 

Figure \ref{fig:5.5}: The expectation value of the magnetization
  calculated from the Landau--Zener formula (solid line) for the 
  case (\protect\ref{eq:param-onemode}) with $ \tilde{t}_{\rm max} =
  500 $. The result is consistent with the numerical data of the
  magnetization if we average the data over the oscillations due to
  spin precession. 

Figure \ref{fig:5.6}: The energy spectrum of the Hamiltonian with two
  boson modes: Case (i) $ \tilde{\omega}_1 = 0.3 $ and 
  $ \tilde{\omega}_2 =  0.35 $. The abscissa is the magnetic field 
  $ \tilde{H} $.

Figure \ref{fig:5.7}: The time evolution of the magnetization 
  $ \langle \sigma^z \rangle_{\tilde{t}} $ (dotted line) and the total
  energy $ \tilde{E}(\tilde{t}) $ (dashed line) in the case (i) with 
  $ \tilde{v} = 1 / 500 $. The ground state energy for each value of 
  $ \tilde{H} $ is also shown as a solid line.

Figure \ref{fig:5.8}: The time evolution of the probabilities $ p_i $
  in the case (i) with $ \tilde{v} = 1 / 500 $.

Figure \ref{fig:5.9}: The energy spectrum of the Hamiltonian with two
  boson modes: Case (ii) $ \tilde{\omega}_1 = 0.3 $ and 
  $ \tilde{\omega}_2 = 0.31 $. 

Figure \ref{fig:5.10}: The time evolution of the eigenstate weights 
  $ p_i $ in the case (ii) with $ \tilde{v} = 1 / 500 $.

Figure \ref{fig:5.11}: The time evolution of the magnetization 
  $ \langle \sigma^z \rangle_{\tilde{t}} $ (dotted line) and the total
  energy $ \tilde{E}(\tilde{t}) $ (dashed line) in the case (ii) with 
  $ \tilde{v} = 1 / 500 $. The ground state energy for each value of 
  $ \tilde{H} $ is also shown as a solid line.

Figure \ref{fig:5.11.5}: The estimates of the effective gaps 
  $ \Delta \tilde{E}  ^{({\rm eff})} $ (solid  line), 
  $ \Delta \tilde{E}_2^{({\rm eff})} $ (dashed line), and 
  $ \Delta \tilde{E}_3^{({\rm eff})} $ (dotted line) for different
  values of $ \tilde{v} $. The estimates barely depend on 
  $ \tilde{v} $.

Figure \ref{fig:5.12}: The energy spectrum of the Hamiltonian with two
  boson modes: Case~(iii) $ \omega_1 = \omega_2 = 0.3 $.

Figure \ref{fig:5.13}: The time evolution of the eigenstate weights 
  $ p_i $ in the case (iii) with $ \tilde{v} = 1 / 500 $. The weights 
  $ p_3 $, $ p_5 $, $ p_6 $, $ p_8 $, and $ p_9 $ vanish for all 
  $ \tilde{t} $.

Figure \ref{fig:5.14}: The time evolution of the magnetization 
  $ \langle \sigma^z \rangle_{\tilde{t}} $ (dotted line) and the total
  energy (dashed line) in the case (iii) with $ \tilde{v} = 1 / 500 $.
  The ground state energy for each value of $ \tilde{H} $ is also
  shown as a solid line. 

Figure \ref{fig:5.15}: The $ \tilde{\omega}_2 $-dependence of the
  effective gaps. The difference between the solid line and the dashed
  line is $ [ \Delta \tilde{E}^{({\rm eff})}_1 ]^2 $ and that between
  the dashed line and the dotted line is 
  $ [ \Delta \tilde{E}^{({\rm eff})}_2 ]^2 $. The combined energy gap
  for the two scattering points around $ \tilde{H} \sim 0.3 $ is given
  by Eq.\ (\ref{eq:gap-combined}), or the the dashed line.

Figure \ref{fig:5.17}: The effective gap for the scattering points
  around $ \tilde{H} = 0.3 $ and the real energy gaps near those
  scattering points. The solid line denotes the combined effective gap 
  $ (\Delta \tilde{E}^{({\rm eff})})^2 $, the dashed line denotes the
  left-hand side of Eq.\ (\ref{eq:eff-case1}), 
  $ (\Delta \tilde{E}_2)^2 + (\Delta \tilde{E}_3)^2 $,
  and the dotted line denotes 
  $ (\Delta \tilde{E}_2    +  \Delta \tilde{E}_3)^2 $.

\newpage

\begin{figure}
\caption{A schematic of the energy spectrum of the two-level system}
\label{fig:5.0}
\end{figure}

\begin{figure}
\caption{The energy spectrum in the one-mode case 
  (\protect\ref{eq:param-onemode}). The abscissa is the magnetic field 
  $ \tilde{H} $. The labels $ | \downarrow , 0 \rangle $ and
  $ | \uparrow , n \rangle $ indicate the relevant physical
  interpretation of the states, where $ n $ is the number of excited
  bosons.} 
\label{fig:5.1}
\end{figure}

\begin{figure}
\caption{The time evolution of the magnetization $ \langle \sigma^z
  \rangle_{\tilde{t}}$ (dotted line) and the total energy 
  $ \tilde{E}(\tilde{t}) $ (dashed line). The
  field was changed as $ \tilde{H} = (1/500) \tilde{t}$. 
  The ground-state energy
  for each value of $ \tilde{H} $ is also shown as the solid line.}
\label{fig:5.2}
\end{figure}

\begin{figure}
\caption{The time evolution of the magnetization 
  $ \langle \sigma^z \rangle_{\tilde{t}}$ (dashed line) and the energy
  of the spin, 
  $\tilde{E}_{\rm spin} (\tilde{t})
  \equiv \langle - \frac{1}{2} \tilde{\delta} \sigma^x 
  - \frac{1}{2} \tilde{H}(\tilde{t}) \sigma^z 
    \rangle_{\tilde{t}}$ (solid line).} 
\label{fig:5.3}
\end{figure}

\begin{figure}
\caption{The time evolution of the eigenstate weights $ p_i $. The solid
  line $ p_1 $ indicates the component of the ground state.}
\label{fig:5.4}
\end{figure}

\begin{figure}
\caption{The expectation value of the magnetization calculated from
  the Landau--Zener formula (solid line) for the 
  case~(\protect\ref{eq:param-onemode}) with $ \tilde{t}_{\rm max} =
  500 $. The    result
  is consistent with the numerical data of the magnetization 
  if we average the data over the 
  oscillations due to spin precession.}
\label{fig:5.5}
\end{figure}

\begin{figure}
\caption{The energy spectrum of the Hamiltonian with two boson
  modes: Case (i) $ \tilde{\omega}_1 = 0.3 $ and 
  $ \tilde{\omega}_2 =  0.35 $. 
  The abscissa is the magnetic field $ \tilde{H} $.}
\label{fig:5.6}
\end{figure}

\begin{figure}
\caption{The time evolution of the magnetization $ \langle \sigma^z
  \rangle_{\tilde{t}} $ (dotted line) and the total energy
  $\tilde{E}(\tilde{t})$  
  (dashed line) in the case~(i) with $\tilde{v}=1/500$. 
  The ground state energy 
  for each value of $ \tilde{H} $ is also shown as a solid line.} 
\label{fig:5.7}
\end{figure}

\begin{figure}
\caption{The time evolution of the probabilities $ p_i $ in the case (i) 
with $ \tilde{v} = 1 / 500 $.}
\label{fig:5.8}
\end{figure}

\begin{figure}
\caption{The energy spectrum of the Hamiltonian with two boson modes: 
  Case (ii) $ \tilde{\omega}_1 = 0.3 $ and $ \tilde{\omega}_2 = 0.31 $.}
\label{fig:5.9}
\end{figure}

\begin{figure}
\caption{The time evolution of the eigenstate weights $ p_i $ in the 
case (ii) with $ \tilde{v} = 1 / 500 $.}
\label{fig:5.10}
\end{figure}

\begin{figure}
\caption{The time evolution of the magnetization $ \langle \sigma^z
  \rangle_{\tilde{t}} $ (dotted line) and the total energy
  $\tilde{E}(\tilde{t})$  
  (dashed line) in the case~(ii) with $ \tilde{v} = 1 / 500 $.
  The ground state energy 
  for each value of $ \tilde{H} $ is also shown as a solid line.}
\label{fig:5.11}
\end{figure}

\begin{figure}
\caption{The estimates of the effective gaps 
$\Delta \tilde{E}^{({\rm eff})}$,
$\Delta \tilde{E}_{2}^{({\rm eff})}$ and 
$\Delta \tilde{E}_{3}^{({\rm eff})}$
for different values of $\tilde{v}$.
The estimates barely depend on $\tilde{v}$.}
\label{fig:5.11.5}
\end{figure}

\begin{figure}
\caption{The energy spectrum of the Hamiltonian with two boson modes: 
Case~(iii)
  $ \omega_1 = \omega_2 = 0.3 $.}
\label{fig:5.12}
\end{figure}

\begin{figure}
\caption{The time evolution of the eigenstate weights $ p_i $ in the 
case (iii) with $ \tilde{v} = 1 / 500 $. The weights $ p_3 $, $ p_5 $, 
  $ p_6 $, $ p_8 $, and $ p_9 $ vanish for all $ \tilde{t} $.}
\label{fig:5.13}
\end{figure}

\begin{figure}
\caption{The time evolution of the magnetization $ \langle \sigma^z
  \rangle_{\tilde{t}} $ (dotted line) and the total energy (dashed line)
   in the case~(iii) with $\tilde{v}=1/500$.
  The ground state energy 
  for each value of $ \tilde{H} $ is also shown as a solid line.} 
\label{fig:5.14}
\end{figure}

\begin{figure}
\caption{The $ \tilde{\omega}_2 $-dependence of the effective gaps.}
\label{fig:5.15}
\end{figure}

\begin{figure}
\caption{The effective gap for the scattering points around
  $\tilde{H}=0.3$
  and the real energy gaps near those scattering points.}
\label{fig:5.17}
\end{figure}


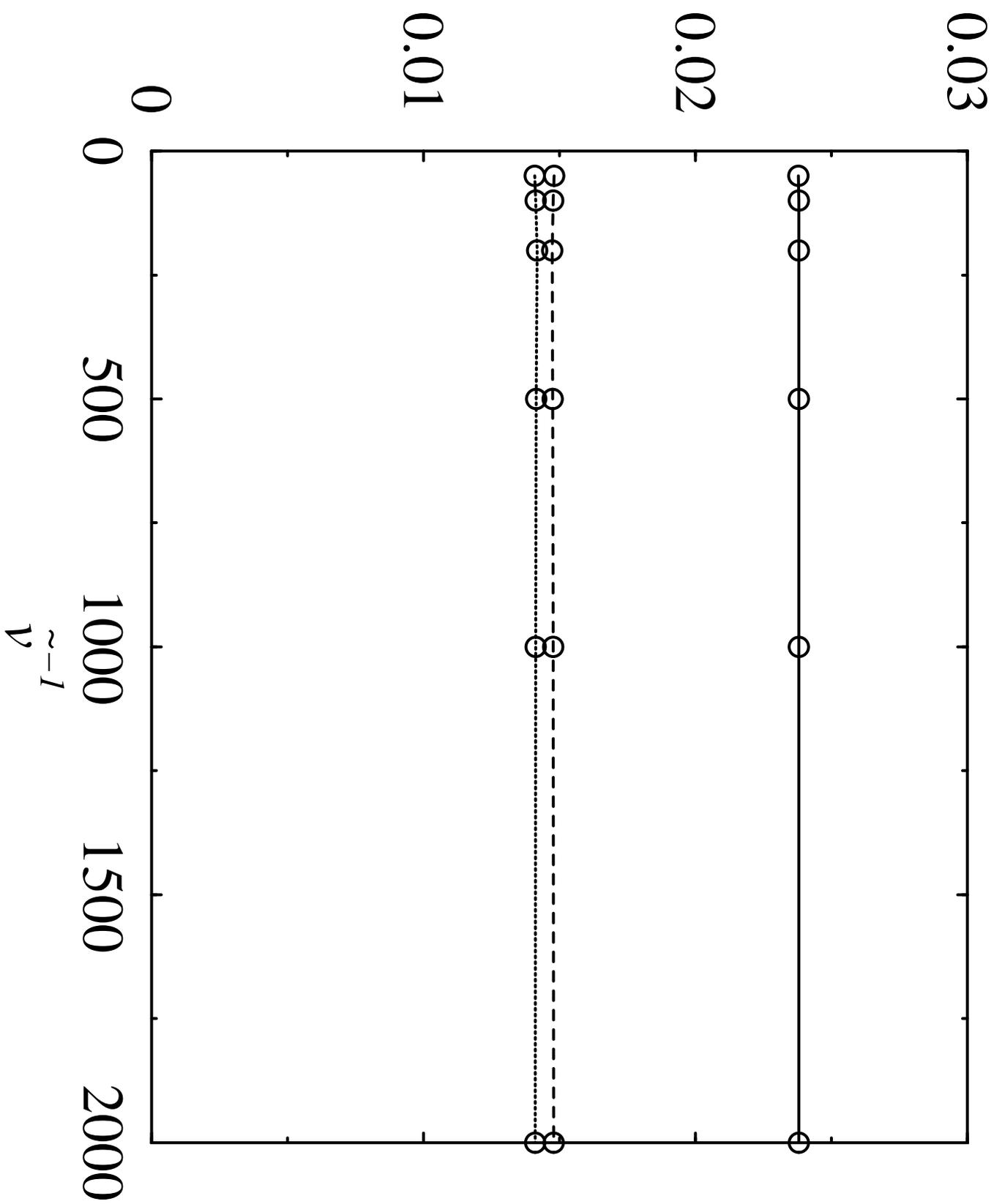
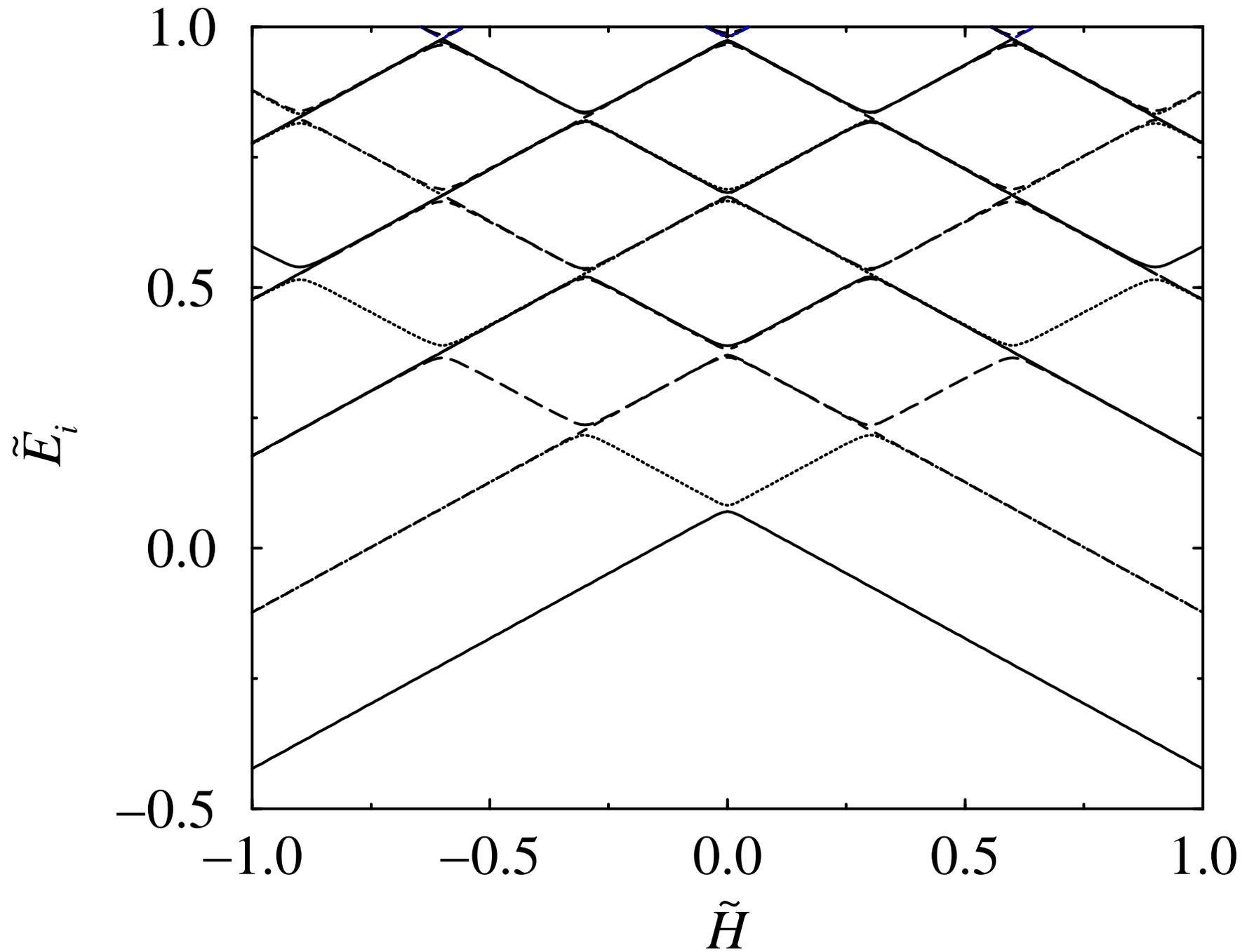
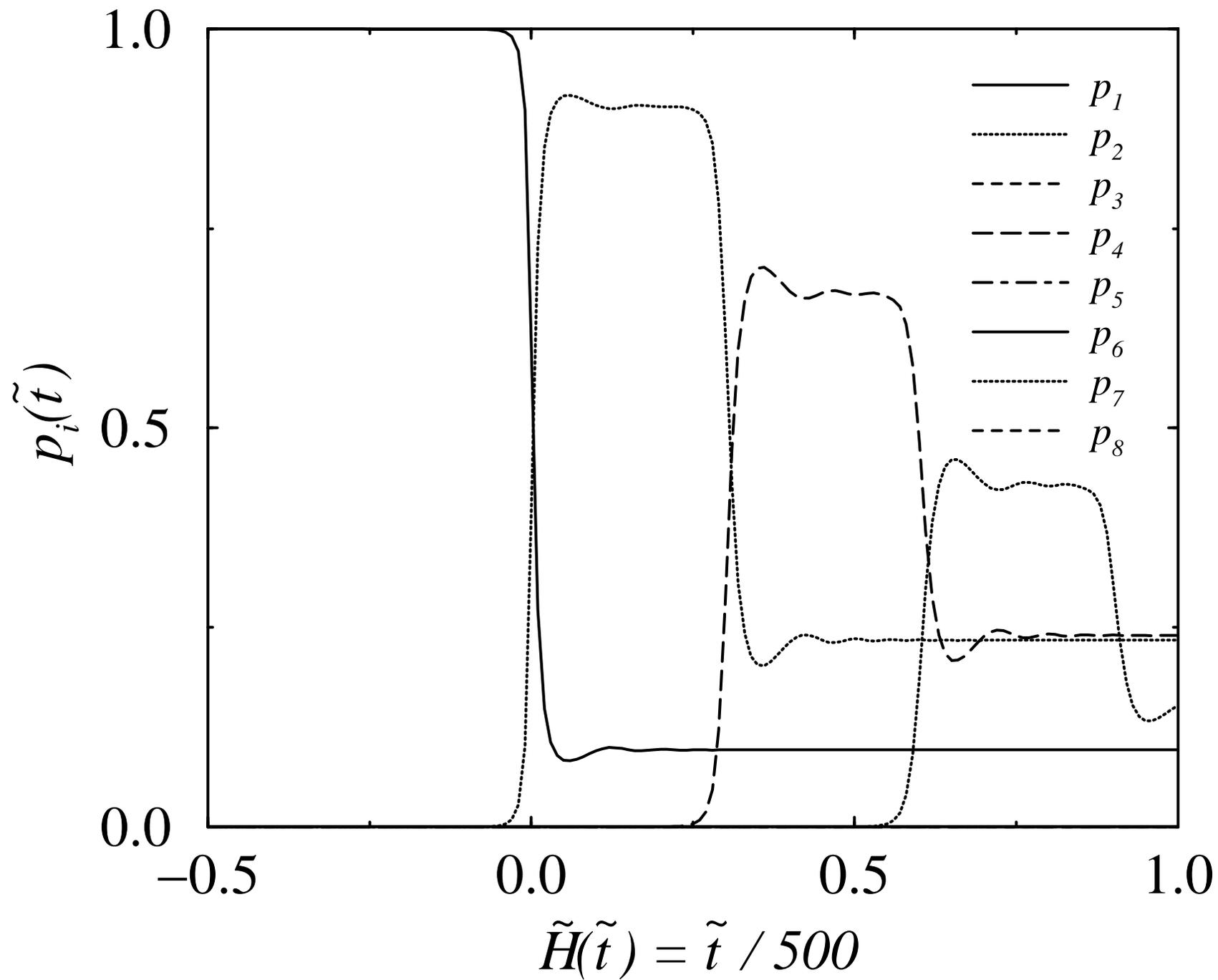
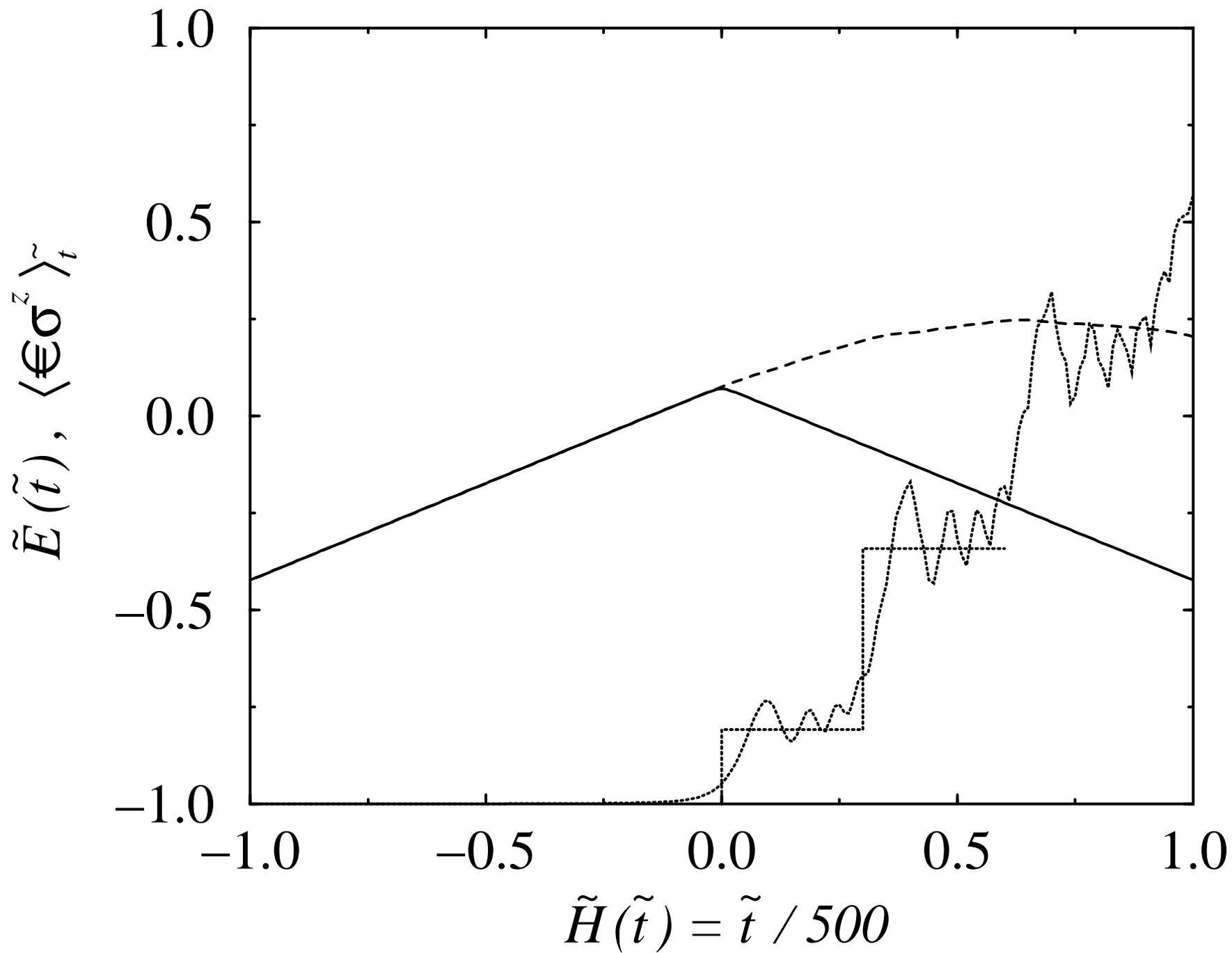
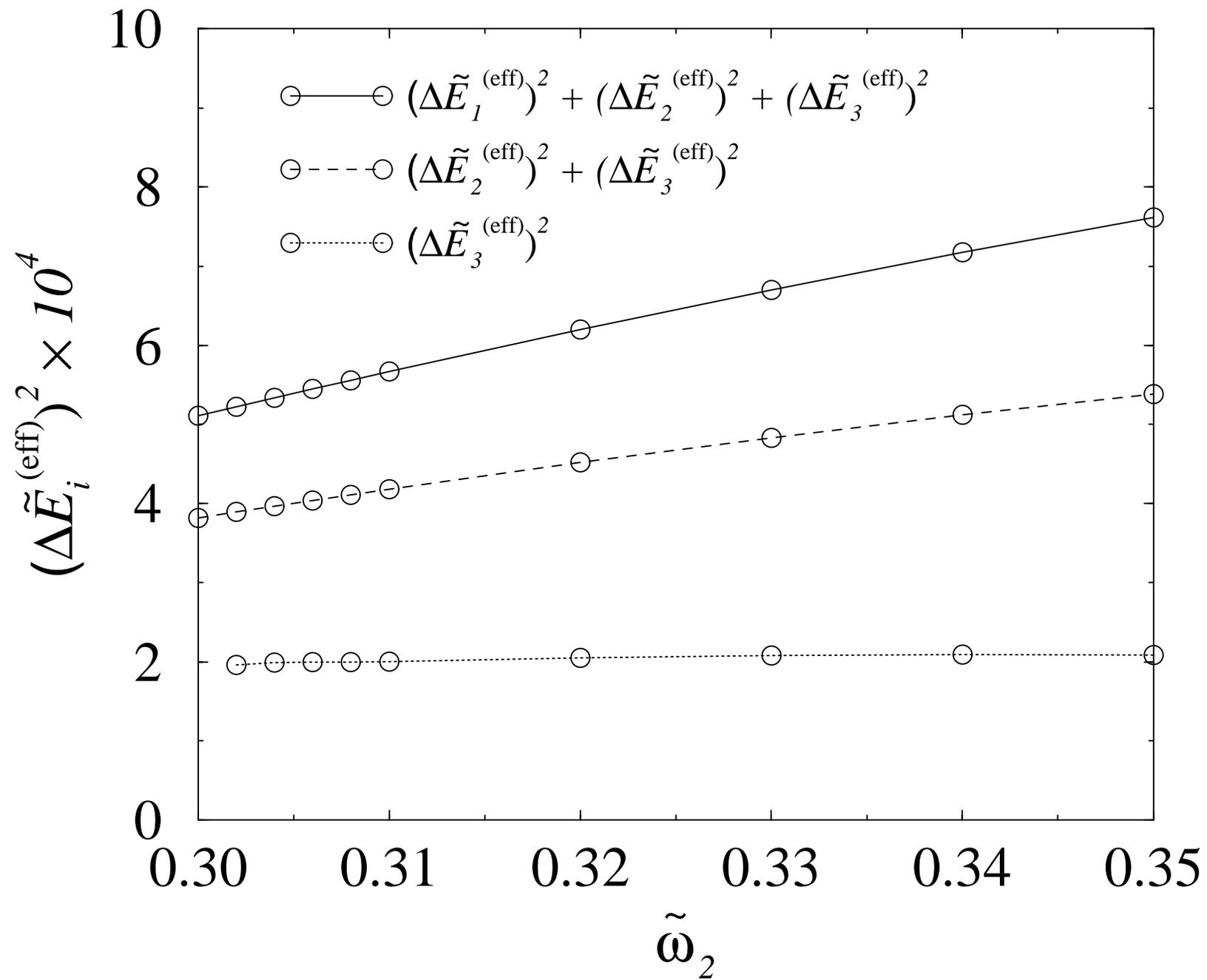
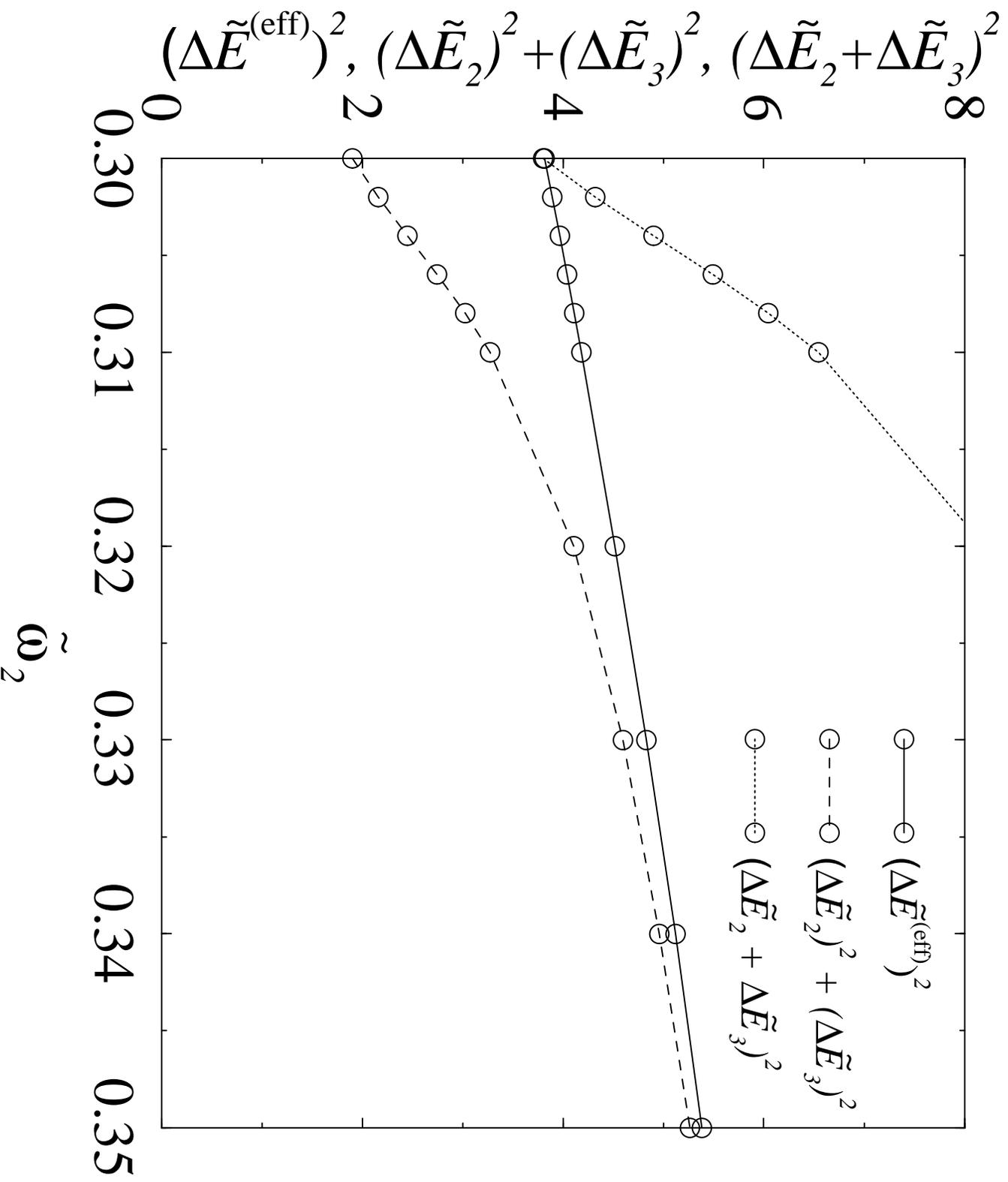
\end{document}